\documentclass[aps,showpacs,superscriptaddress,twocolumn,prl]{revtex4-1}
\usepackage{amssymb}
\usepackage{natbib}
\usepackage{amsmath}
\usepackage{amsfonts}
\usepackage{graphicx}
\usepackage{mathrsfs}
\usepackage{dcolumn}
\usepackage{bm}
\usepackage{color}
\usepackage{cancel}
\usepackage{mathbbol}
\usepackage{epsfig}
\usepackage{units}
\usepackage{esint}
\usepackage{soul} 

\definecolor{rot}{rgb}{0.75,0.05,0.25}
\definecolor{hellgrau}{gray}{0.5}
\definecolor{blau}{rgb}{0,0,0.7}

\def\Tr{\mbox{Tr}}

\begin{document}

\textbf{Comment on ``Experimental Verification of a Jarzynski-Related Information-Theoretic Equality by a Single Trapped Ion''  PRL \textbf{120} 010601 (2018)
}

Reference \cite{Xiong18PRL120} reports on the experimental verification of an identity in probability  theory that reads (see Eq. (3) of Ref. \cite{Xiong18PRL120}):
\begin{align}
\langle e^{-I_{nm}}\rangle = \sum_{mn} p_{nm}e^{-I_{nm}} =1
\label{eq:AvExpI}
\end{align}
It is claimed in Ref. \cite{Xiong18PRL120} that Eq. (\ref{eq:AvExpI}) implies Eq. (\ref{eq:JI}) below via the relation (see Eq. (4) of Ref. \cite{Xiong18PRL120}):
\begin{align}
I_{nm} = \beta (E'_m-E_n- F'+F)
\qquad \text{(not correct)}
\label{eq:wrong}
\end{align}
Our comment is that while (under specific conditions) Eqs. (\ref{eq:AvExpI}, \ref{eq:JI}) can be simultaneously valid, it is not true that the quantities appearing in the exponents are the same. Equation (\ref{eq:wrong}) is not valid. Accordingly, it is not correct to state that Eq. (\ref{eq:AvExpI}) is related to, or implies Eq. (\ref{eq:JI}) \footnote{We are not using the notation $\Delta F=F-F'$ employed in Ref.  \cite{Xiong18PRL120}, which has opposite sign convention as compared to the standard notation \cite{Campisi11RMP83}, in order to minimise confusion. We also refrain from using the notation $W=E_n-E'_m$ of \cite{Xiong18PRL120} which would hint at the fact that the energy change in the system can be interpreted as work. This is in fact generally not the case, in particular it is not the case for open quantum systems (a situation that can in certain cases be described by CPTP maps), in which case the change in a system energy generally contains both work and heat \cite{Talkner09JSM09,Campisi11RMP83}.}.

Equation (\ref{eq:AvExpI}) presents an identity that holds for any joint probability $p_{nm}$. It follows from the Bayes rule $p_{nm}=p_{m|n}p_n$, normalisation  $\sum_{nm}p_{nm}= \sum_{n} p_n =\sum_m q_m=1$ and the definition $I_{nm}= \ln p_{m|n} -\ln {q_m}$ in \cite{Xiong18PRL120}:
\begin{align}
\sum_{mn} p_{nm}e^{-I_{nm}} = \sum_{mn} \frac{ p_{nm} q_m}{p_{m|n}}=  \sum_{n} p_n \sum_m q_m=1 \, .
\label{eq:proof}
\end{align}

Equation (\ref{eq:wrong}) is claimed to be valid ``if the system is initially prepared in the thermal Gibbs state''\cite{Xiong18PRL120}. 
We first point out that  if that were true, upon inserting Eq. (\ref{eq:wrong}) in Eq. (\ref{eq:AvExpI}), it would imply that \begin{align}
\langle e^{-\beta (E'_m-E_n- F'+F)}\rangle= 1
\label{eq:JI}
\end{align}
for all quantum systems evolving according to a CPTP map, starting in thermal equilibrium and being subject to a two-point energy measurement, for which $p_{nm} = \Tr\, Q_m \sum_i \Lambda_i (P_n \rho P_n) \Lambda_i^\dagger Q_m$ \cite{Xiong18PRL120}.
This contradicts the well known fact that, under those conditions, rather the following holds true \cite{Morikuni11JSP143,Kafri12PRA86,Rastegin13JSTAT13,Albash13PRE88,Watanabe14PRE89,Campisi17NJP19}:
\begin{align}
\langle e^{-\beta (E'_m-E_n- F'+F)}\rangle= \gamma = \sum_i \Tr\, \Lambda_i^\dagger \rho \Lambda_i
\end{align}
with $\gamma$ being generally different from $1$
\footnote{A condition for $\gamma$ to equal 1 is that the dynamics be unitary, or more generally, unital \cite{Morikuni11JSP143,Kafri12PRA86,Rastegin13JSTAT13,Albash13PRE88,Watanabe14PRE89,Campisi17NJP19}; i.e.  $\sum_i \Lambda_i\Lambda^{\dagger}_i = \mathbb 1$. }.

That Eq. (\ref{eq:wrong}) is not valid can be checked by direct inspection. For $I_{nm}$ we find
\begin{align}
I_{nm}&= \ln \frac{p_{m|n}}{q_m}=  \ln \frac{p_{nm}}{p_n q_m}= \ln \frac{\Tr\, Q_m \sum_i \Lambda_i P_n \rho P_n \Lambda_i^\dagger}{(\Tr\, P_n \rho) (\Tr\, Q_m \sum_i \Lambda_i \rho  \Lambda_i^\dagger) } \nonumber \\
&= \ln \frac{(\sum_{k}e^{-\beta E_k})\,  (\Tr\,  Q_m \sum_i \Lambda_i P_n \Lambda_i^\dagger)}{\Tr\, Q_m \sum_{i,k} e^{-\beta E_k} \Lambda_i  P_k  \Lambda_i^\dagger }
\label{eq:Inm-expand}
\end{align}
where we have used $\sum_i\Lambda_i^\dagger \Lambda_i= \mathbb{1}$,  $\rho= e^{-\beta H}/Z= \sum_k P_k e^{-\beta E_k}/Z$, $P_k P_n = \delta_{kn}P_n$, $ q_m=\sum_n p_{nm}=\Tr\,  Q_m \sum_i \Lambda_i P_n \Lambda_i^\dagger e^{-\beta E_n}/Z$ and $p_n= \sum_m p_{nm}= \Tr\, P_n \rho= e^{-\beta E_n}/Z$ \cite{Xiong18PRL120}.
For $\beta (E'_m-E_n-F'+F)$
we find
\begin{align}
\beta (E'_m-E_n-F'+F)= \ln \frac{(\sum_k e^{-\beta E'_k}) e^{\beta (E'_m - E_n)} }{ \sum_k e^{-\beta E_k}}
\label{eq:Wnm-expand}
\end{align}
Note that the final eigenvalues $E'_k$ enter explicitly in Eq. (\ref{eq:Wnm-expand})  while Eq.  (\ref{eq:Inm-expand}) is {\it independent} of the final eigenvalues $E'_k$. Similarly, the projectors $P_k, Q_m$ enter Eq. (\ref{eq:Inm-expand}) explicitly whereas they do not appear in Eq. (\ref{eq:Wnm-expand}). Therefore the two quantities are never equal. This argument remains valid as well for the special case of an unitary evolution, $U_C$, that commutes with the  $P_k$'s, as employed in Ref. \cite{Xiong18PRL120}, for which we obtain
\begin{align}
I_{nm} = \ln \frac{(\sum_{k}e^{-\beta E_k})\,  (\Tr\,  Q_m P_n)}{\Tr\, Q_m \sum_{k} e^{-\beta E_k}P_k   }
\label{eq:Inm-U}
\end{align}
In that case Eq. (\ref{eq:AvExpI}) and Eq. (\ref{eq:JI}) hold simultaneously  (as verified experimentally in Ref. \cite{Xiong18PRL120}) but Eq. (\ref{eq:wrong}) is,  however,  nevertheless {\it not} valid.
\vspace{5mm}
\newline
Michele Campisi$^1$ and Peter H\"anggi$^2$
\newline
\small{$^1$Dipartimento di Fisica e Astronomia, Universit\`a di Firenze and INFN Sezione di Firenze, Via G. Sansone 1, I-50019 Sesto Fiorentino (FI), Italy
\newline
$^2$Institute of Physics, University of Augsburg, Universit\"atsstra\ss e 1, D-86135 Augsburg, Germany}

\end{document}